\documentclass[%
amsmath,amssymb,
reprint,
]{revtex4-1}
\usepackage{graphicx}
\usepackage{dcolumn}
\usepackage{bm}

\usepackage{amsmath}
\usepackage{color}
\usepackage[normalem]{ulem}
\usepackage{balance}
\usepackage[T1]{fontenc}

\makeatletter
\renewcommand*{\@fnsymbol}[1]{\ifcase#1\or*\else\@arabic{#1}\fi}
\makeatother

\usepackage{xcolor}
\usepackage{xpatch}
   \makeatletter
   \def\changeBibColor#1{%
     \in@{#1}{} %
     \ifin@\color{red}\else\normalcolor\fi
   }
   \xpatchcmd\@bibitem
   {\item}
   {\changeBibColor{#1}\item}
   {}{}
  
   \xpatchcmd\@lbibitem
   {\item}
   {\changeBibColor{#2}\item}
   {}{}

   \makeatother

\begin{document}
\title{Extend the random-walk shielding-potential viscosity model to hot temperature regime}

\author{Yuqing Cheng}\affiliation{School of Mathematics and Physics, University of Science and Technology Beijing, Beijing 100083, China}%
\author{Xingyu Gao}\affiliation{Laboratory of Computational Physics, Institute of Applied Physics and Computational Mathematics, Beijing 100094, China}%
\author{Qiong Li}\affiliation{Laboratory of Computational Physics, Institute of Applied Physics and Computational Mathematics, Beijing 100094, China}%
\author{Yu Liu}\affiliation{Laboratory of Computational Physics, Institute of Applied Physics and Computational Mathematics, Beijing 100094, China}%
\author{Haifeng Song}\affiliation{Laboratory of Computational Physics, Institute of Applied Physics and Computational Mathematics, Beijing 100094, China}%
\author{Haifeng Liu}\thanks{Corresponding author: liu\_haifeng@iapcm.ac.cn}
\affiliation{Laboratory of Computational Physics, Institute of Applied Physics and Computational Mathematics, Beijing 100094, China}%

\begin{abstract}
\textbf{ABSTRACT: } The transport properties of matter have been widely investigated. In particular, shear viscosity over a wide parameter space is crucial for various applications, such as designing inertial confinement fusion (ICF) targets and determining the Rayleigh-Taylor instability.
In this work, an extended random-walk shielding-potential viscosity model (ext-RWSP-VM) based on the statistics of random-walk ions and the Debye shielding effect is proposed to elevate the temperature limit of RWSP-VM [Phys. Rev. E 106, 014142] in evaluating the shear viscosity of one component plasma.
In the extended model, we reconsider the collision distance that is introduced by hard-sphere concept, hence, it is applicable in wide temperature regime rather than a narrower one in which RWSP-VM is applicable.
The results of H, C, Al, Fe, Ge, W, and U show that the extended model provides a systematic way to calculate the shear viscosity of arbitrary one component plasma at the densities from about 0.1 to 10 times the normal density (the density at room temperature and 1 standard atmosphere). This work will help to develop viscosity model in wide regime when combined with our previous low temperature viscosity model [AIP Adv. 11, 015043].
\end{abstract}
\maketitle

\section{\label{sec:Introduction}Introduction}
Warm dense matter (WDM) and hot dense matter (HDM) usually refer to the temperature regime from several electronvolts (eV) to thousands of eV. The transport properties, especially the shear viscosities of WDM and HDM are significant in numerous applications, such as designing inertial confinement fusion (ICF) targets \cite{appICF1,appICF2}, determining the Rayleigh-Taylor instability \cite{appRT1,appRT2}, understanding the evolution of astrophysical objects \cite{appAstro1,appAstro2}, understanding wave damping in dense plasmas \cite{appDamping}, microjetting and stability analysis of interface during shock loading \cite{appMjet,Stable1,Stable2}.

Since the experiments of viscosity measurements are very difficult for WDM and HDM, researchers have been developing several theoretical methods to obtain demanded viscosity data.
For example, first-principles molecular dynamics (FPMD) simulations \cite{methodsFPMD} combined with the Green-Kubo relations \cite{methodGK} are first used by Alf\'e and Gillan to calculate the viscosity of liquid Al and Fe-S.
Hou $et$ $al$. employ both classical molecular dynamics (CMD) and Langevin molecular dynamics (LMD) simulations which are based on average-atom model combined with the hyper-netted chain approximation (AAHNC) to obtain the viscosities of metals (Al, Fe, and U) \cite{HouAl,RVM,HouU}.
The molecular dynamics (MD) simulations are accurate, and the accuracy is ensured by the number of Kohn-Sham orbitals \cite{KS} which increase rapidly when the temperature increases. It results in the high cost of MD simulations, which makes it difficult to obtain sufficient amount of data.
Therefore, developing more efficient viscosity model is necessary.
Stanton-Murillo transport model (SMT) \cite{SMT1} is based on the Boltzmann equation and employs the Coulomb logarithm to evaluated the momentum-transfer cross section in the Boltzmann equation, and it can be used to calculate the viscosity for high-energy-density matter.
In the one-component plasma model (OCP), Daligault $et$ $al$. \cite{OCP} use the equilibrium MD simulations from the weakly coupled regime to the solidification threshold and a practical expression is proposed from the fitting of the MD data to calculate the viscosity.
Cl\'erouin $et$ $al$. extend the Wallenborn-Baus formula \cite{extOCP} by using the data calculated from MD simulations to form an extension of the Wallenborn-Baus fit, and obtain the extended Wallenborn-Baus (EWB) formula, \cite{Clerouin} which is applicable to an arbitrary mixture.
However, for OCP and EWB, the parameter space of the MD simulations would limit their applicable regime.
Therefore, in a previous work \cite{RVM}, we developed an analytical model called ``random-walk shielding-potential viscosity model'' (RWSP-VM) to evaluate the shear viscosity. It is based on the statistics of random-walk ions and the Debye shielding effect. Compared with the models and the simulation methods mentioned above, it is applicable in warm temperature regime, and it has the advantages of universality, accuracy, and efficiency.

In this work, we extend RWSP-VM by reconsidering the collision distance which is introduced by the hard-sphere concept. This new model named as ext-RWSP-VM has a wider temperature range of applications, i.e., it is applicable in the temperature regime of $10^0-10^5$ eV, maintaining the advantages of RWSP-VM simultaneously.
For simplicity, we roughly denote the temperature regime of $10^0-10^2$ eV and $10^2-10^5$ eV as warm temperature regime and hot temperature regime, respectively.

\section{\label{sec:Model}Model}
\begin{figure}[tb]
\includegraphics[width=0.48\textwidth]{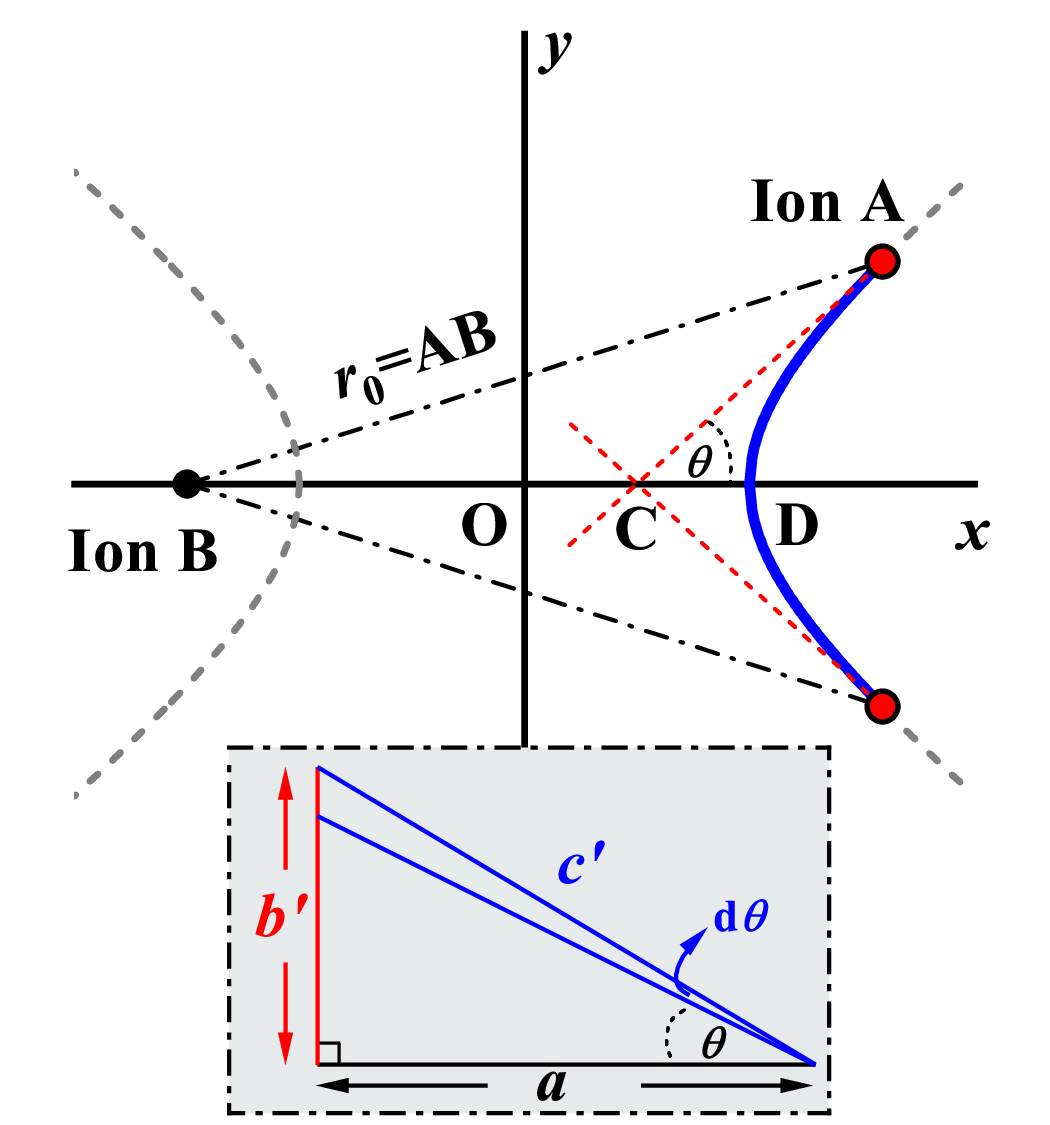}
\caption{\label{fig:scheme} Schematic of the hyperbolic track of the ions. The blue curve stands for the hyperbolic track of Ion A in the reference frame of Ion B. The red dashed curves stand for the tangent of the hyperbola at point A and its symmetric point about the x-axis. $r_0$ is the cutoff distance, i.e., there is interaction between the two ions only when the separation distance $r$ is smaller than $r_0$. $\theta$ is the incident angle between the incident velocity and $x$-axis. The inset is the scheme of the integration over $\theta$, where $b^{\prime}=\frac{R}{\sqrt{R^2-c^2}}~b$ and $c^{\prime}=\sqrt{\frac{R^2-a^2}{R^2-c^2}}~c$.
}
\end{figure}

As the temperature increases, the ionization of the metal increases. Particularly, in warm temperature regime, electrons are partially ionized; while in hot temperature regime, most or all of the electrons are ionized. It results in the fact that the metal ions are principally influenced by the Coulomb interactions between each other. The interactions led to the motion variation of the ions, which can be treated as a kind of ``collisions'', which is called as the Coulomb collisions or distant collisions due to the long range nature of the Coulomb potential. On the other hand, in both regime, the ion velocities are high enough that they can be treated as ``ion gas'' which move randomly, so that the concept of random-walk can process their behaviors properly. Moreover, the Debye shielding effect plays an important role, and the real interactions among these ions are the shielding Coulomb potential, which can be processed by considering the Coulomb potential with a cutoff distance $r_0$, i.e., the Coulomb potential works only if $r<r_0$.
Therefore, the main part of the track of the ions is hyperbola, as shown in Fig. \ref{fig:scheme}. Here, the hyperbolic equation is $\frac{x^2}{a^2}-\frac{y^2}{b^2}=1$, where $a$ is the semi-major axis and $b$ is the semi-minor axis, with the focal length $c=\sqrt{a^2+b^2}$.
The viscosity $\eta$ is evaluated by the formula below \cite{RVM}:
\begin{equation}
\eta=\frac{\sqrt{3 m k_B T}}{\pi d^4} I.
\label{eq:eta}
\end{equation}
The derivation of this formula can be found in both Ref. \cite{RVM} and the Appendix of this work.
Here $m$ is the atom mass, $k_B$ is the Boltzmann constant, $T$ is the temperature, $d$ is the collision distance introduced by hard-sphere concept, and $I$ is a quantity that is relevant to $T$, which can be expressed as:
\begin{equation}
I=2r_0^2 \frac{K\left[ 2 ( 1-K ) + ( 1+K )\mathrm{ln}K \right]}{( 1+\sqrt{K} )^2( 1-K )^2}.
\label{eq:IT}
\end{equation}
Here, $K=\left( \frac{R}{a} \right)^2$, $R\equiv r_0 -a$, $a=\frac{q^2}{3 k_B T+2 q^2 / r_0}$, $q^2=\frac{(\overline{Z} e)^2}{4 \pi \varepsilon_0}$, $\overline{Z}=\frac{1}{n}\sum\limits_{j=0}^Z j n_j$ is the average ionization, $Z$ is the nuclear charge, $n$ is the total number density of the ions, $n_j$ is the number density of the $j$th ionized ion, and $r_0$ is the cutoff distance which is introduced by the Debye shielding effect. In RWSP-VM, we assumed that both $d$ and $r_0$ equal to the Debye length $\lambda_D\equiv \sqrt{\frac{\varepsilon_0 k_B T}{n_e e^2 (z^*+1)}}$, where $z^* \equiv \overline{Z^2}/\overline{Z}$ and $\overline{Z^2}=\frac{1}{n}\sum\limits_{j=0}^Z j^2 n_j$.
The approximation $d=\lambda_D$ is successful in warm temperature regime. However, in hot temperature regime, the large $\lambda_D$ and the almost full-ionization both make the ions interact more strongly; moreover, the speed of the ions are so large that the collision distance is much smaller than the cutoff distance, resulting in the failure of $d=\lambda_D$, thus the inapplicability of RWSP-VM. Because $r_0$ and $d$ represent the maximum and the minimum distances between two ions within interaction, respectively, and the higher $T$ is, the smaller $d$ is; while $\lambda_D$ increases with the increasing $T$. When $T$ is high, $d$ is much smaller than $\lambda_D$, resulting in the fact that $d\neq \lambda_D$.
Therefore, the limit of RWSP-VM is that it is applied only in warm temperature regime though it has good characteristics such as universality, accuracy, and high efficiency. In this work, we extend it to hot temperature regime by carefully reconsidering $d$ analytically, i.e., calculating its average value over scattering angles of the collisions between ions. The extended model is called ext-RWSP-VM.

Here, the assumption that the cutoff distance is equal to the Debye length is still a good approximation:
\begin{equation}
r_0=\lambda_D.
\label{eq:r0}
\end{equation}
However, we should carefully reconsider $d$. The track of the ions is hyperbola, as shown in Fig. \ref{fig:scheme}.  The minimum distance between two ions should be $(a+c)$. For certain temperature $T$, $a$ is determined, but $b$ varies with the scattering angles of the collisions, resulting in the varying $c$. Therefore, we calculate the average value of $d$ over incident angles:
\begin{equation}
\begin{aligned}
d&=\overline{a+c}=\frac{\int_{0}^{\theta_\mathrm{m}}(a+c)\sin \theta ~ \mathrm{d}\theta}{\int_{0}^{\theta_\mathrm{m}} \sin \theta ~ \mathrm{d}\theta} \\
&=a\left[ 1+\frac{R}{b_\mathrm{m}}\mathrm{ln}(\frac{R+b_\mathrm{m}}{a}) \right].
\end{aligned}
\label{eq:d}
\end{equation}
Here, $\theta_\mathrm{m} =\frac{\pi}{2}$, $\sin \theta = \frac{b^{\prime}}{c^{\prime}} =\frac{b}{b_\mathrm{m}}\frac{R}{c}$, and $b_\mathrm{m} = \sqrt{R^2-a^2}$. The relationships among these parameters are displayed in Ref. \cite{RVM} and can be interpreted in the inset of Fig. \ref{fig:scheme}, e.g., $b$ ranges from 0 to $b_\mathrm{m}$ which results in $b^{\prime}$ ranging from 0 to $+\infty$ and $\theta$ ranging from 0 to $\frac{\pi}{2}$. 
Moreover, $b^{\prime}$ and $c^{\prime}$ are induced to make $\theta$ and the integration in Eq. (\ref{eq:d}) clearer in mathematics. In physics, the scattering angle between the two ions in one collision equals to $\pi-2\theta$. Hence, small $\theta$ is corresponding to large scattering angle, and vice versa. In Eq. (\ref{eq:d}), $(a+c)\sin \theta$ tells the probability of the scattering, indicating that larger scattering angle has smaller scattering probability, which agrees with the opinion in Chapter 11 of Ref. \cite{PlasmaPhysics}.
We emphasize that the distance of closest approach between ions varies with the incident angle $\theta$. When calculating $d$, all the incident angles are considered. Hence, ext-RWSP-VM not only employs the hard-sphere concept, but also is beyond the hard-sphere concept.
Combining Eqs. (\ref{eq:eta})-(\ref{eq:d}), we obtain the analytical expressions of the shear viscosity for ext-RWSP-VM. 

We have compared several methods which calculate $\overline{Z}$ (Thomas-Fermi (TF) model \cite{TF1,TF2}, density functional theory MD \cite{DFT}, path integral Monte Carlo approach \cite{PIMC}, AAHNC approximation \cite{AAHNC1}, and the Hartree-Fock-Slater model \cite{HFS}, etc.) in Ref. \cite{RVM}, and concluded that all of these methods agree well with each other with certain differences in warm and hot temperature regime. Due to the simplicity, efficiency, and accuracy of TF model \cite{TF1,TF2}, we choose it to calculate the average ionization $\overline{Z}$.
In general, $z^*=\overline{Z}$ is a good approximation unless otherwise specified.

\section{\label{sec:Results}Results and Discussions}
\begin{figure}[tb]
\includegraphics[width=0.48\textwidth]{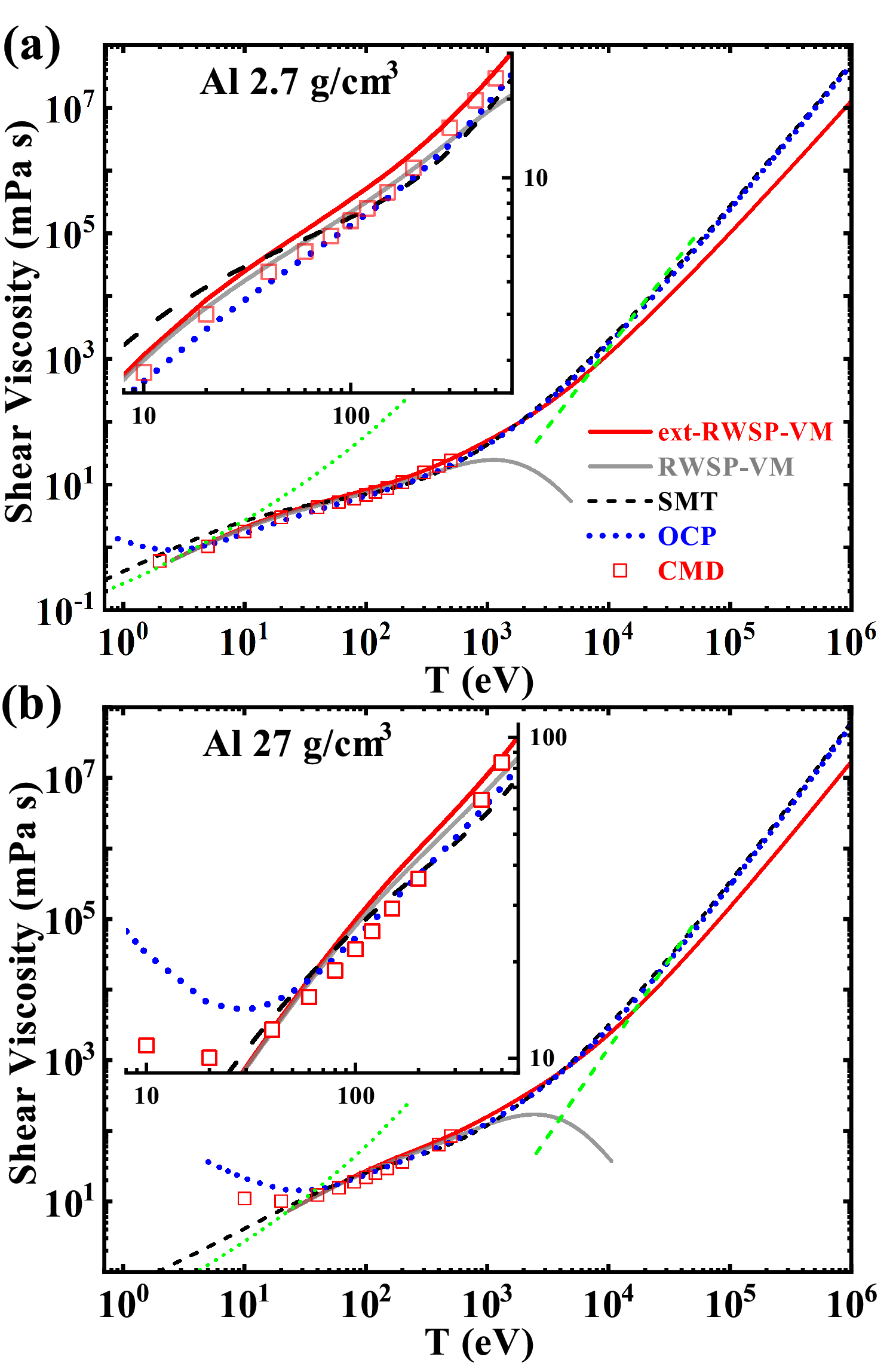}
\caption{\label{fig:Al0} Shear viscosity of Al at the density of 2.7 $\mathrm{g/cm^3}$ (a) and 27 $\mathrm{g/cm^3}$ (b).
Red solid, gray solid, black dashed, and blue dotted curves stand for the results of ext-RWSP-VM, RWSP-VM \cite{RVM}, SMT \cite{SMT1,SMT2}, and OCP \cite{OCP}, respectively. Open squares stand for the results of CMD \cite{HouAl}.
The green dotted curves represent the lower limits of both RWSP-VM and ext-RWSP-VM, and the green dashed curves represent the upper limits of ext-RWSP-VM.
Each inset stands for the zoom of temperature from about 10 to 600 eV.
}
\end{figure}
First, we take Al as an example to illustrate the validity of ext-RWSP-VM. Fig. \ref{fig:Al0} shows the shear viscosity of Al at the densities of 2.7 and 27 $\mathrm{g/cm^3}$. The results of ext-RWSP-VM are compared with the ones of RWSP-VM \cite{RVM},  SMT \cite{SMT1,SMT2}, OCP \cite{OCP}, and CMD \cite{HouAl}.
Here, the applicable regime of each model is clarified in their articles. SMT is applicable for $\Gamma<10$, $\Gamma<20$, and $\Gamma<100$ for $\kappa=1$, $\kappa=2$, and $\kappa=3$, respectively. The expression of OCP is applicable for $\Gamma<200$ in the case of $\kappa=0$.
Here, $\Gamma=q^2/(a_{ws}k_B T)$ is the coupling parameter, $a_{ws}=(\frac{3}{4 \pi n})^{1/3}$ is the Wigner-Seitz radius of the ion, and $\kappa$ is the screening parameter.

In the temperature regime of about 10-500 eV and at the densities of both 2.7 and 27 $\mathrm{g/cm^3}$, ext-RWSP-VM agrees well with RWSP-VM, and the former is slightly larger than the latter. Both of them agree well with OCP, SMT, and CMD. In the temperature regime of about over 500 eV and at the densities of both 2.7 and 27 $\mathrm{g/cm^3}$, RWSP-VM fails, but the trend of ext-RWSP-VM agree with SMT and OCP with some underestimation at about $T>10^4$ eV. At the density of 8.1 $\mathrm{g/cm^3}$, the phenomenon (which is not shown in Fig. \ref{fig:Al0}) is almost similar to the ones of 2.7 and 27 $\mathrm{g/cm^3}$.
The extended model corrects the collision distance $d$ of the previous model. In the previous model, $d$ equals to the cut-off distance $r_0$, which, however, is not proper when the temperature is high enough. Because $d$ should be the closest distance between the ions, the difference between $d$ and $r_0$ become large at high temperature. Therefore, in the extended model, $d$ is estimated by its average value over the scattering angles of the collisions between ions. 
It can be concluded that ext-RWSP-VM is applicable in wide temperature regime, i.e., $T$ satisfies $\lambda_D(T)>0.1a_{ws}$ \cite{RVM}. In the approximation of $z^*+1 \approx \overline{Z}$ for large $\overline{Z}$, the relation can be easily reduced to $\Gamma<100/3 \approx 33.3$. In other words, the lower temperature limit is estimated $T_{\mathrm{lower}}$ by $\Gamma(T_{\mathrm{lower}}) \approx 33.3$.

In the insets of Fig. \ref{fig:Al0}, we notice that ext-RWSP-VM behaves similarly as SMT does when compared with CMD. For temperature above $10^4$ eV, SMT and OCP agree well with each other, but they both present larger viscosity than ext-RWSP-VM does.
Hence, ext-RWSP-VM also has its upper temperature limit, which is shown as the green dashed curves in Fig. \ref{fig:Al0}. The details will be discussed in the next paragraph. 
\begin{figure}[tb]
\includegraphics[width=0.48\textwidth]{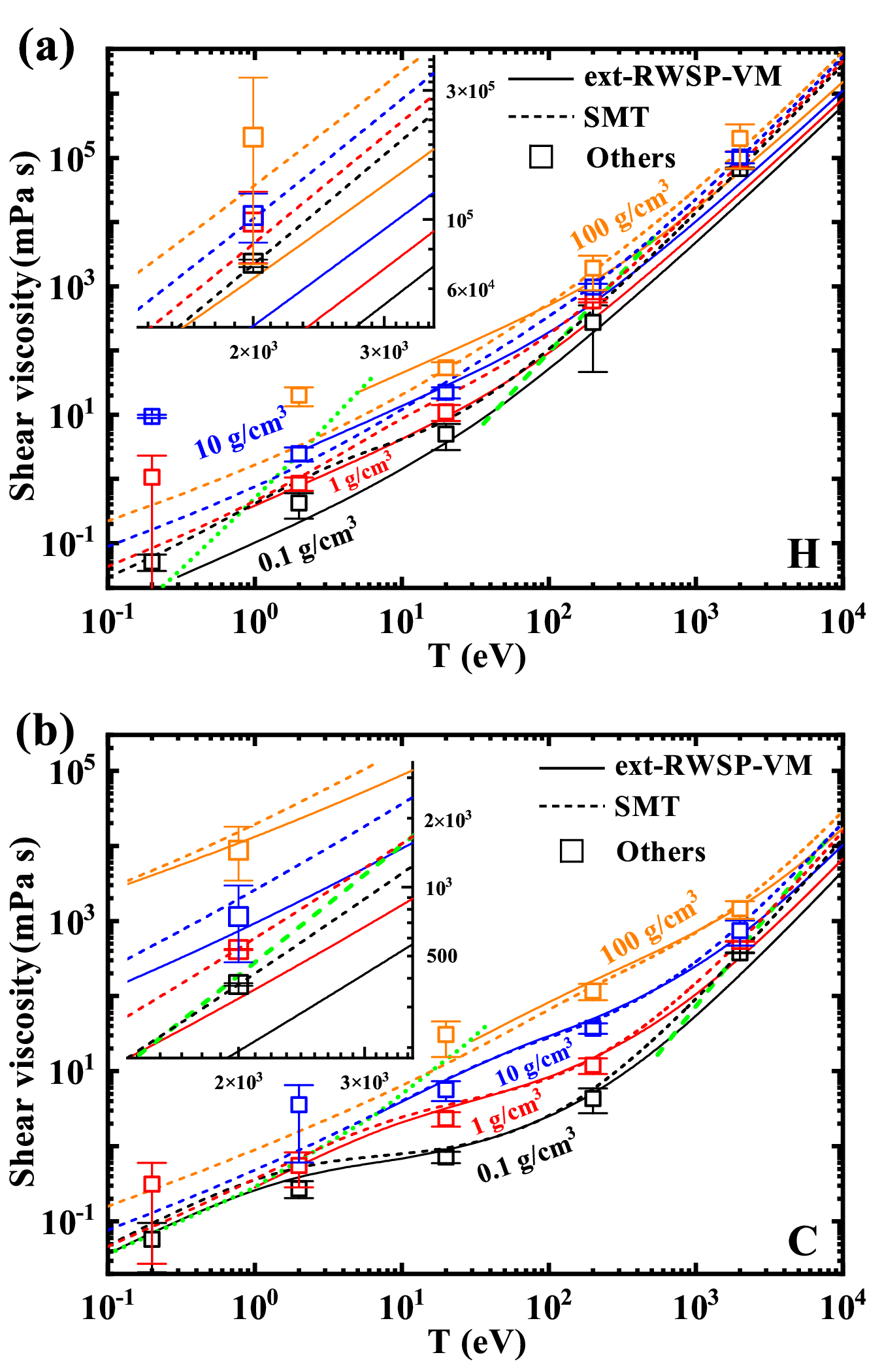}
\caption{\label{fig:EtaHC} Shear viscosity of H (a) and C (b).
Black, red, blue, and orange stand for the densities of 0.1, 1, 10, and 100 $\mathrm{g/cm^3}$, respectively. Solid curves, dashed curves, and open squares, stand for the results of ext-RWSP-VM, SMT \cite{SMT1,SMT2}, and the average of several methods \cite{HCdata} calculated from several methods, respectively.
The green dotted and dashed curves represent the lower and upper limits of the temperature regime, respectively in which ext-RWSP-VM is applicable.
Each inset stands for the zoom of temperature around 2000 eV.
}
\end{figure}

Second, we take H and C as examples to explore the upper temperature limit of our model.
Fig. \ref{fig:EtaHC} shows the shear viscosity of both H and C at densities from 0.1 to 100 $\mathrm{g/cm^3}$ and at temperature from 0.2 to 2000 eV.
We compare the results of ext-RWSP-VM with the ones of SMT \cite{SMT1,SMT2} and the statistical average of several methods \cite{HCdata} including the pseudo-ion in jellium (PIJ) model \cite{PIJ}, the effective potential theory with average atom model (EPT+AA) \cite{EPT1,EPT2}, OFMD \cite{OFMD2}, the hybrid kinetics-molecular dynamics (KMD) \cite{KMD}, and AAHNC \cite{AAHNC2}. 

For H, generally, our model agrees with the MD methods when the temperature is less than 2000 eV. In particular, at $T=200$ eV, our model agrees well with them at the density of 100 $\mathrm{g/cm^3}$, but not well at lower densities; at $T=2000$ eV, SMT agrees well with them, but our models underestimate the viscosity.

For C, generally, our model agrees with the MD methods. In particular, at $T=2000$ eV, our model agrees well with them at the densities of both 10 and 100 $\mathrm{g/cm^3}$, but differs from them at lower densities.

It indicates that the upper temperature limit varies with the density. We can also use the coupling parameter $\Gamma$ to express. According to the analysis above, we calculate $\Gamma$ of H and C at conditions that are close to the possible application range limits. That is, $\Gamma=0.18$ at $\rho=10\ \mathrm{g/cm^3}$ and $T=200$ eV for H; $\Gamma=0.15$ at $\rho=1\ \mathrm{g/cm^3}$ and $T=2000$ eV for C. Therefore, the upper temperature limit is estimated to satisfy $\Gamma(T_{\mathrm{upper}}) \approx 0.2$. In other words according to the upper limit, ext-RWSP-VM satisfies $\Gamma>0.2$. It is obvious that $T_{\mathrm{upper}}$ increases as the density increases.

It is worth mentioning that $z^*=\overline{Z}$ is not a proper approximation for H. For a certain H atom, its net charge is either 1 or 0, i.e., we find a H atom with a charge of 1 and 0 with the probability of $\overline{Z}$ and $ (1-\overline{Z})$, respectively. According to the definition, $\overline{Z^2}=\overline{Z} \times 1^2 + (1-\overline{Z})\times 0^2=\overline{Z}$, hence, $z^* \equiv \overline{Z^2}/\overline{Z}= 1 $.
When consider the collision process, the interaction is related to $q^2$. A more accurate treatment is that only $\overline{Z}$ proportion of the H atoms have charge 1 and be considered for the collisions, i.e., $(1-\overline{Z})$ proportion is ignored. In other words, we do not treat every ion with charge $\overline{Z}$ as we have done for other elements. 
Therefore, when calculating the quantities, such as $q^2$, $a$, $R$, $K$, $b_\mathrm{m}$, and $d$, which are related to $\overline{Z}$, we should use $\overline{Z}=\overline{Z}_{\mathrm{eff}}=1$ instead. According to the above revision, the viscosity of H (solid curves in Fig. \ref{fig:EtaHC}a) is calculated with our model.

For a brief summary, the ext-RWSP-VM model indicates that as the temperature or/and the density increases, the viscosity increases. The applicable range of this model can be evaluated by $0.2 <\Gamma< 33.3$.
\begin{figure}[tb]
\includegraphics[width=0.48\textwidth]{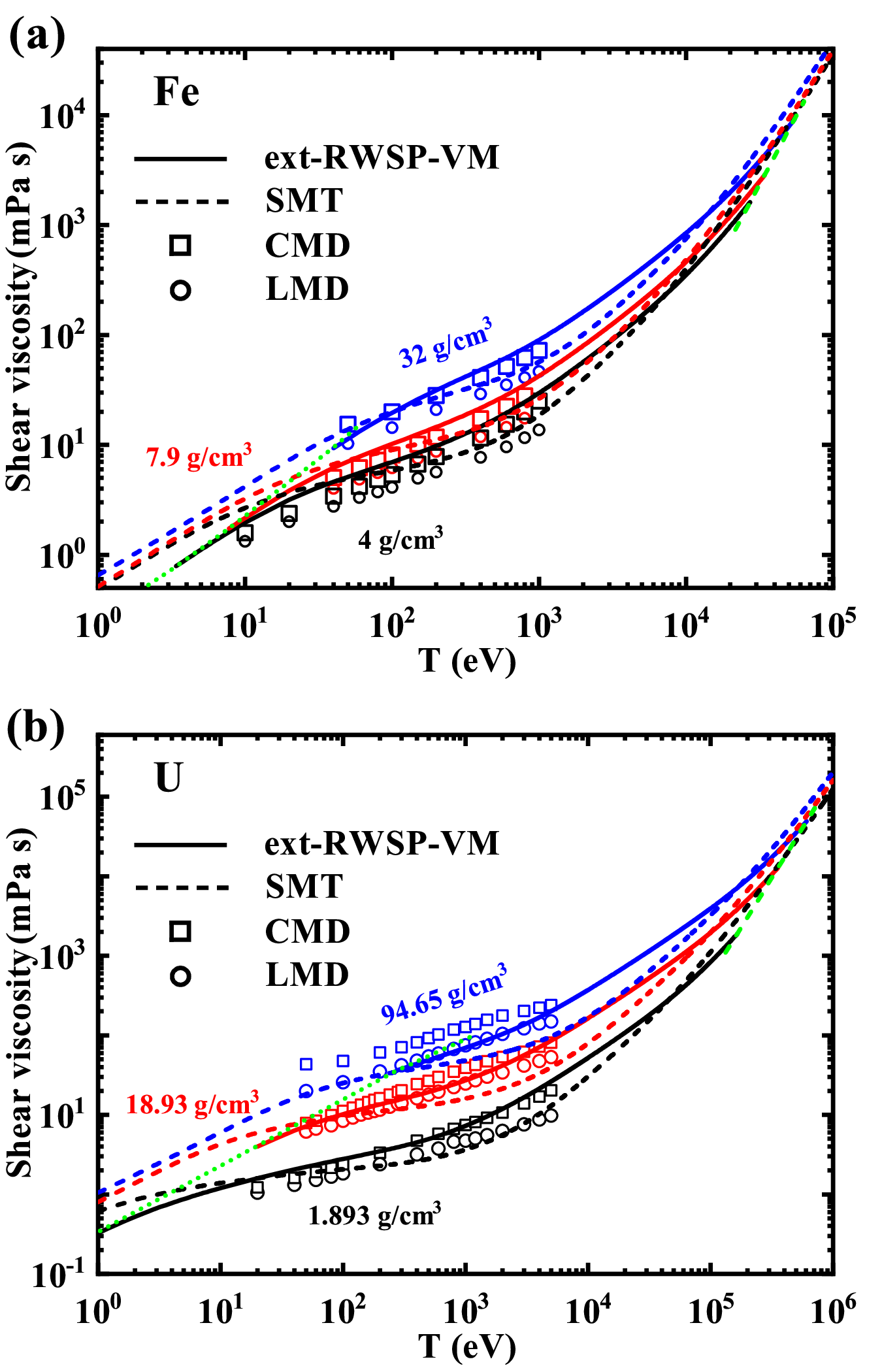}
\caption{\label{fig:EtaFeU} Shear viscosity of Fe (a) and U (b). Solid and dashed curves stand for the results of ext-RWSP-VM, SMT \cite{SMT1,SMT2}, respectively; open squares and circles stand for the ones of CMD and LMD \cite{RVM}, respectively. Black, red, and blue stand for the densities of 4.0, 7.9, and 32 $\mathrm{g/cm^3}$ for Fe (a) respectively, and the densities of 1.893, 18.93, and 94.65 $\mathrm{g/cm^3}$ for U (b), respectively. 
The green dotted and dashed curves represent the lower and upper temperature limits of ext-RWSP-VM.
}
\end{figure}

\begin{figure}[tb]
\includegraphics[width=0.48\textwidth]{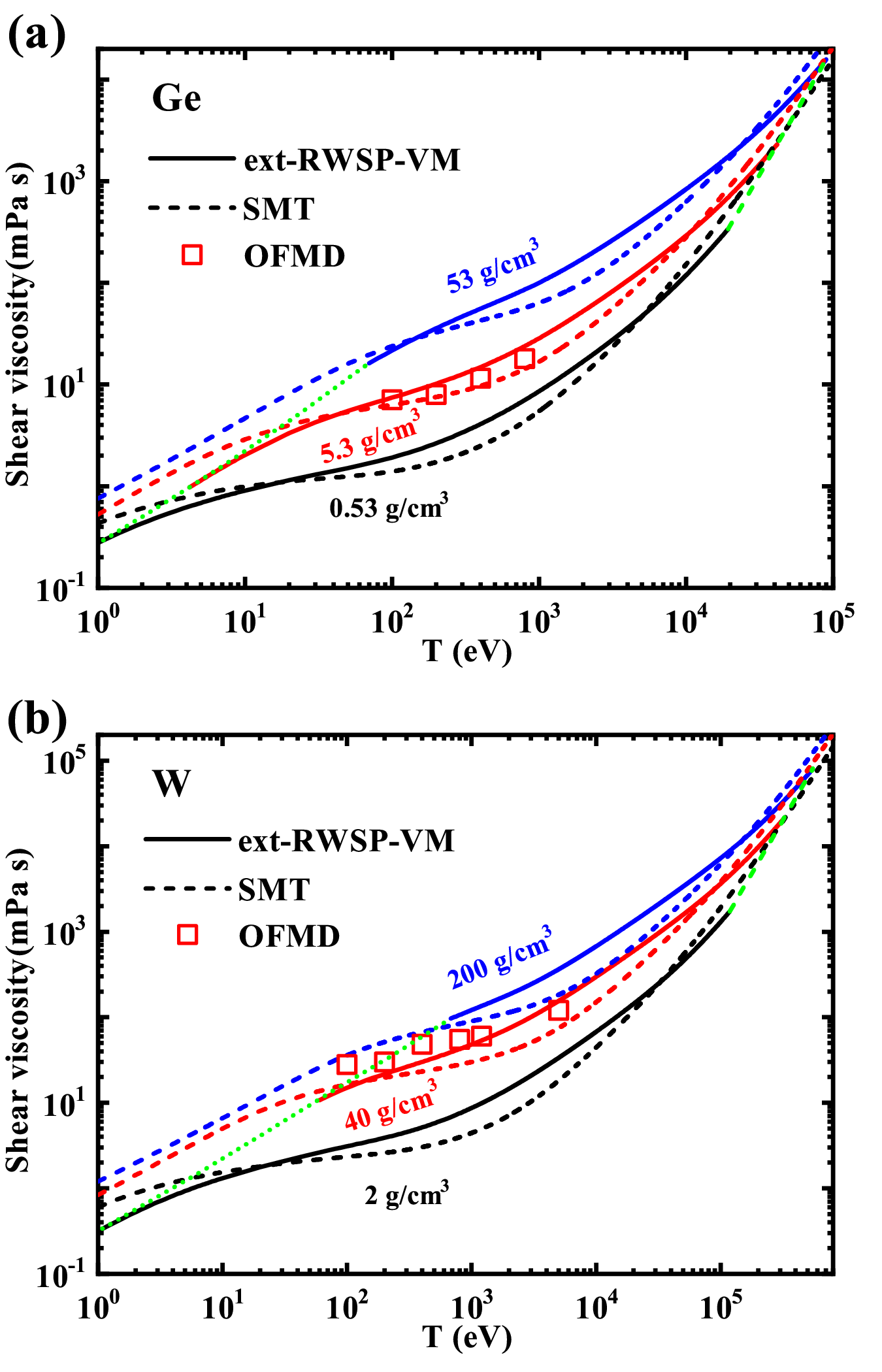}
\caption{\label{fig:EtaWGe} Shear viscosity of Ge (a) and W (b). Solid curves, dashed curves, and open squares stand for the results of ext-RWSP-VM, SMT \cite{SMT1,SMT2}, and OFMD \cite{OFMD}, respectively. For Ge, black, red, and blue stand for the densities of 0.53, 5.3, 53 $\mathrm{g/cm^3}$, respectively. For W, black, red, and blue stand for the densities of 2, 40, 200 $\mathrm{g/cm^3}$, respectively. The green dotted and dashed curves represent the lower and upper temperature limits of ext-RWSP-VM, respectively.
}
\end{figure}
Third, in addition to Al, H, and C, we take several other elements, i.e., Fe, U, Ge, and W, as examples to show the results of ext-RWSP-VM, as shown in Fig. \ref{fig:EtaFeU} and Fig. \ref{fig:EtaWGe}, respectively. The results calculated from ext-RWSP-VM are compared with the ones of the MD simulations and SMT. 

When compared with MD, in general, for all these elements, ext-RWSP-VM agrees with MD simulations. In particular, for Al and Fe, ext-RWSP-VM agrees well with CMD, but for U, ext-RWSP-VM agrees well with CMD at lower density (1.893 $\mathrm{g/cm^3}$) and with Langevin MD simulations (LMD) at higher density (94.65 $\mathrm{g/cm^3}$), the derivations of which have already been discussed in RWSP-VM \cite{RVM}. Here, we summarize in brief that at higher densities, the interaction between ions are weaken by the nonadiabatic dynamic effects. Because for high-Z elements at higher densities, the ionic structures are complex with the increase temperature. In this case, LMD is more appropriate than CMD due to the fact that LMD considers interaction between ions and electrons, which corresponds to the Debye shielding effect in our model. 

Furthermore, ext-RWSP-VM agrees well with Thomas-Fermi orbital-free formulation coupled with molecular dynamics (OFMD) \cite{OFMD,TFMD}. As mentioned  in Ref. \cite{TFMD}, OFMD uses the Thomas-Fermi kinetic energy functional for the electrons to modify scheme of density functional theory, and it is well suited to perform very-high-temperature MD simulations, e.g., 5 keV.
Particularly, in Fig. \ref{fig:EtaWGe}, $T_{\mathrm{upper}}$ of W can reach about $10^5$ eV. It also indicates that $T_{\mathrm{upper}}$ depends on the density of the matter.

When compared with SMT, in general, ext-RWSP-VM presents larger viscosity than SMT does in warm temperature regime, but the situation is reversed in high temperature regime.
Particularly, in the case of Fe, SMT agrees with CMD. However, in the case of U, SMT agrees with LMD at the densities of 1.893 and 18.93 $\mathrm{g/cm^3}$, but underestimates the viscosity at 94.65 $\mathrm{g/cm^3}$.
In the case of Ge, both ext-RWSP-VM and SMT agree well with OFMD at the density of 5.3 $\mathrm{g/cm^3}$. However, in the case of W, ext-RWSP-VM behaves a little better than SMT does at the density of 40 $\mathrm{g/cm^3}$.

Finally, we would illustrate that the approximations in ext-RWSP-VM are reasonable, and present the applicable regime of this model in the view of WDM and HDM.
For WDM, one of the definitions can be described as ``a condensed matter regime characterized roughly by electron temperatures about 1-15 eV and pressures to 1 Mbar or greater'' \cite{FrontiersWDM1}. The approximations we made can be summarized as Coulomb potential with cut-off distance (Debye length) and binary collisions, which we denote as condition A. In general, condition A is suitable in plasma physics which is corresponding to HDM. However, it is not suitable for WDM. Because for WDM, some other effects cannot be ignored, i.e., correlations, quantum degeneracy, and screening etc., which we denote as condition B. It is clear that condition B is not used in our model. 

Although there is not a precise boundary between WDM and HDM, an approximation criterion can be estimated by employing the electron degeneracy parameter $\Theta \equiv \frac{k_B T}{E_f}$, where $E_f \equiv \frac{\hbar^2}{2 m_e} (3 \pi^2 n_e)^{2/3}$ is the Fermi energy \cite{FrontiersWDM2}. When $\Theta>1$, corresponding to HDM, condition B can be ignored; when $\Theta$ is close to 1, corresponding to WDM, condition B begins to play an important role. Therefore, the boundary between WDM and HDM can be roughly estimated as $\Theta(T_\mathrm{b})=1$ so that the corresponding temperature boundary $T_\mathrm{b}$ can be calculated. 

In Table \ref{tab:limit}, we list some $T_\mathrm{lower}$ (defined as $\Gamma(T_\mathrm{lower})=33.3$), $T_\mathrm{b}$, and  $T_\mathrm{upper}$ (defined as $\Gamma(T_\mathrm{upper})=0.2$) for the elements involved in this work at certain densities.
In most cases, they satisfy $T_\mathrm{lower} < T_\mathrm{b} < T_\mathrm{upper}$. This indicates that the applicable regime of ext-RWSP-VM crosses WDM and HDM. Due to the fact that $T_\mathrm{lower}$ is close to $T_\mathrm{b}$ compared with the difference between $T_\mathrm{b}$ and $T_\mathrm{upper}$, we can figure out that our model is also applicable in certain regime below the upper limit of WDM, although our model only consider condition A rather than condition B which is important for WDM.
We also notice that in some cases, i.e., W and U at high densities, $T_\mathrm{b} < T_\mathrm{lower} < T_\mathrm{upper}$. This indicates that the applicable regime of ext-RWSP-VM is for HDM only. Hence, condition A is enough in these cases.
Therefore, our model is applicable in HDM regime and in certain regime below the upper limit of WDM.

\begin{table}[htbp] %
\centering 
\caption{The lower temperature limit $T_\mathrm{lower}$, the boundary temperature $T_\mathrm{b}$, and the upper temperature limit $T_\mathrm{upper}$ as a function of element and its density.}  
\label{tab:limit} 

\begin{tabular}{c | c | c | c | c } 
\hline 
Element & Density ($\mathrm{g/cm^3}$) & $T_\mathrm{lower}$  (eV) &  $T_\mathrm{b}$ (eV) & $T_\mathrm{upper}$ (eV)\\ 
\hline 
	& 0.1	& 0.26	& 3.7	& 37.3	\\
H	& 1.0	& 0.81	& 22.0	& 82.6	\\
	& 10.0	& 2.1	& 112.1	& 182.4	\\	
\hline 
 	& 0.2 	& 0.19	& 1.6	& 875.7		\\
C 	& 2.0 	& 1.8	& 13.9  & 1.892$\times 10^3$	\\
 	& 20.0 	& 10.7	& 91.8  & 4.082$\times 10^3$	\\
\hline 
	& 0.27	& 0.18	& 1.2	& 3.524$\times 10^3$	\\ 
Al 	& 2.7	& 2.5	& 11.9 	& 7.600$\times 10^3$	\\
	& 27.0	& 23.0	& 92.3	& 1.638$\times 10^4$	\\
\hline 
	& 0.79	& 0.41	& 2.1	& 1.593$\times 10^4$	\\
Fe	& 7.9	& 7.7	& 21.5	& 3.433$\times 10^4$	\\
	& 79.0	& 88.4	& 174.3	& 7.397$\times 10^4$	\\
\hline 
 	& 0.53	& 0.21	& 1.1  	& 1.937$\times 10^4$  \\
Ge 	& 5.3	& 4.2	& 12.0  & 4.174$\times 10^4$  \\
 	& 53.0	& 65.7	& 108.5 & 8.995$\times 10^4$ \\
\hline
  	& 2.0	& 0.60	& 2.1   & 1.187$\times 10^5$  \\
W  	& 20.0	& 17.8	& 23.6  & 2.557$\times 10^5$ \\
  	& 200.0	& 6.737$\times 10^2$	& 223.0 & 5.510$\times 10^5$ \\
\hline
	& 2.0	& 0.49	& 1.7 	& 1.684$\times 10^5$  \\
U	& 20.0	& 15.5	& 19.4  & 3.628$\times 10^5$  \\
	& 200.0	& 1.606$\times 10^3$& 191.1 & 7.816$\times 10^5$  \\
\hline 
\end{tabular}

\end{table}

From the above elaborations, we can conclude that ext-RWSP-VM is accurate for arbitrary one component plasma and is applicable in both warm and hot temperature regime, and in certain density regime.

\section*{\label{sec:Conclusion}Conclusions}
In conclusion, we develop a viscosity model called ext-RWSP-VM, which is extended from a previous model (RWSP-VM). We employ ext-RWSP-VM to calculate the shear viscosity of a series of elements including H, C, Al, Fe, Ge, W, and U. By comparing the results of our model with the ones of other methods including AAHNC, PIJ, EPT, OFMD, KMD, SMT, and OCP, we show that our model is universal, accurate, and efficient. The applicable regime of our model is extended, i.e., in wide temperature regime of about $10^0-10^5$ eV, and at the densities from about 0.1 to 10 times the normal density (the density at room temperature and 1 standard atmosphere). More accurately, ext-RWSP-VM can be applied in the regime of $0.2 < \Gamma < 33.3$.

The limitation of the current ext-RWSP-VM is that it is not applicable at higher temperature, at which more complex physics should be considered, e.g., ion-electron and electron-electron interactions, etc. The current model needs to be improved. In the future, we will working on extending our model not only to higher temperature but also to be applicable for mixtures. 

\section*{\label{sec:Acknowldegment}Acknowledgment}
We thank the anonymous reviewers for their careful work and thoughtful suggestions that have helped improve this paper substantially.
This work was supported by Laboratory  of  Computational  Physics (Grant No. 6142A05220303).

\section*{Author Disclosures}
The authors have no conflicts to disclose.

\section*{Data availability}
The data and the codes which support the findings of this study are available from the corresponding author upon reasonable request.

\appendix
\section*{\label{Appendix}Appendix: Derivation}
Most of the physical quantities represented by the symbols can be found in Table \ref{tabA:parameters}.
\begin{table}[htbp] %
\centering 
\caption{The physical quantities represented by the symbols}  
\label{tabA:parameters}  
\begin{tabular}{l | l  } 
\hline 
Symbol & Physical quantity  \\ 
\hline 
$\eta$	& shear viscosity\\
\hline 
$a$		& semi-major axis of the hyperbola\\
$b$		& semi-minor axis of the hyperbola\\
$b_\mathrm{m}$	& the maximum of $b$ ($b_\mathrm{m}=\sqrt{R^2-a^2}$)\\
$c$		& focal distance of the hyperbola\\
$d$		& collision distance\\
$k_0$	& slope of the hyperbola at point A\\
$k_B$	& Boltzmann constant \\
$m$  	& mass of the ion  \\
$n$		& total number density of ions\\
$N$		& total number of ions\\
$N_s$	& total collisions of an ion within $t$ ($N_s \approx \nu t$ )\\
$p_{j\beta}$ 	& momentum in the $\beta$ direction of the $j$th ion  \\
$r_0$	& cut-off distance\\
$r_D$		& curvature radius at point D ($r_D=\frac{b^2}{a}$)\\
$r_{j\alpha}$	& coordinate in the $\alpha$ direction of the $j$th ion    \\
$R$		& $r_0-a$	\\
$t$		& time \\ 
$T$		& temperature \\
$v$		& velocity of an ion \\
$v_A$, $v_D$ & velocity of the ion at point A and D, respectively\\
$v_{jy}$ 	& velocity of the $j$th ion in the $y$ direction\\
$V$ 	& volume of the system \\ 
$x$		& coordinate in the $x$ direction of an ion\\
$(x_0,y_0)$ & coordinate of point A\\
$x_j$		& coordinate in the $x$ direction of the $j$th ion\\
$\alpha$, $\beta$ & directions in the Cartesian coordinate system  \\
$\Delta x$	& total change in $x$ of an ion within $t$\\
$\Delta x_k$ & change in $x$ of the $k$th collision of an ion\\
$\Delta v_{ky}$ & change in $v_y$ of the $k$th collision of an ion\\
$\Delta v_y$ & total change in $v_y$ of an ion within $t$\\

$\theta$	& angle between incident velocity and $x$-axis\\
$\lambda_\mathrm{m}$ & mean free path of the ions\\
$\nu$	& collision frequency \\
$<\cdots>$	& ensemble average  \\
\hline 
\end{tabular}

\end{table}

We start from one of the definitions of the shear viscosity, which can be found on Page 60 of Ref. \cite{methodGK}:
\begin{equation}
\eta=\frac{1}{2t}\frac{V}{k_B T} \left< \left( \mathcal{L}_{\alpha \beta}(t)-\mathcal{L}_{\alpha \beta}(0) \right)^2 \right>.
\label{eqA:definition}
\end{equation}
Here, the parameter $\mathcal{L}_{\alpha \beta}=\frac{1}{V}\sum\limits_{j=1}^N r_{j\alpha} p_{j\beta}$ with $\alpha \neq \beta$,  For simplicity, we set $\alpha=x$, $\beta=y$, and $\mathcal{L}_{\alpha \beta}(0)=0$, thus $r_{j\alpha}=x_j$ and $p_{j\beta}=mv_{jy}$. Hence, 
\begin{equation}
\left< \left( \mathcal{L}_{\alpha \beta}(t) \right)^2 \right> = \frac{m^2}{V^2}\sum\limits_{j=1}^N \left<(x_{j} v_{jy})^2 \right> = \frac{m^2}{V^2} N \left< (x v_y)^2 \right>.
\label{eqA:L}
\end{equation}
Within time $t$, there are $N_s \approx \nu t$ ($N_s \gg 1$) collisions for an ion. Therefore, every (the $k$th) collision makes a bit contribution to $x$ and $v_y$, i.e., $\Delta x_k$ and $\Delta v_{ky}$. We obtain:
\begin{equation}
(x v_y)^2 = \left[ \left(\sum\limits_{k=1}^{N_s} \Delta x_k \right) \left(\sum\limits_{k=1}^{N_s} \Delta v_{ky} \right) \right]^2.
\label{eqA:xvy1}
\end{equation}
It is obvious that the right hand of Eq. (\ref{eqA:xvy1}) can be expressed as the sum of some square terms and cross terms. When calculating the ensemble average, the cross terms can be ignored due to the fact that positive and negative values occurs symmetrically with equal probability. Hence, 
\begin{equation}
\label{eqA:xvy2}
\begin{aligned}
\left< (x v_y)^2 \right> 
&	=\left< \left(\sum\limits_{k=1}^{N_s} \Delta x_k \right)^2 \left(\sum\limits_{k=1}^{N_s} \Delta v_{ky} \right)^2 \right>\\
&	=\left< \left( \sum\limits_{k=1}^{N_s}(\Delta x_k)^2 \right)
\left( \sum\limits_{k=1}^{N_s} (\Delta v_{ky})^2 \right)\right> \\
&	=N_s \left<(\Delta x)^2\right>
\left<(\Delta v_y)^2\right>\\
&=N_s \frac{2}{3}\lambda_\mathrm{m}^2 \left<(\Delta v_y)^2\right>.
\end{aligned}
\end{equation}
Here, we use the relation $\left<(\Delta x)^2\right>= \frac{2}{3}\lambda_\mathrm{m}^2$. The relation between $\nu$ and $\lambda_\mathrm{m}$ is:
\begin{equation}
\nu=\frac{v}{\lambda_\mathrm{m}}=\sqrt{2}\pi n d^2 v.
\label{eqA:Lm}
\end{equation}
Thus far, the problem is reduced to calculate $\left<(\Delta v_y)^2\right>$.

According to our model, when $r<r_0$, the trajectory the ion is hyperbola, as shown in Fig. \ref{fig:scheme}, and the hyperbolic equation is:
\begin{equation}
\frac{x^2}{a^2}-\frac{y^2}{b^2}=1,\ (x>0,\ a^2+b^2=c^2).
\label{eqA:hyperbola}
\end{equation}
We can write down the equation of conservation of energy at point A and D:
\begin{equation}
\frac{1}{2}m v_A^2 + \frac{q^2}{r_0}=\frac{1}{2} m v_D^2 + \frac{q^2}{a+c},
\label{eqA:Energy}
\end{equation}
and Newton's second law for ion A at point D:
\begin{equation}
\frac{q^2}{(a+c)^2}=m\frac{v_D^2}{r_D}.
\label{eqA:Newton}
\end{equation}
Here, $r_D=\frac{b^2}{a}$ is the curvature radius of the hyperbola at point D, and $v_A=v$. At temperature $T$, the velocity $v$ satisfies $\frac{1}{2}mv^2=\frac{3}{2}k_B T$ as:
\begin{equation}
v=\sqrt{\frac{3k_B T}{m}}.
\label{eqA:v}
\end{equation}
Then we obtain:
\begin{equation}
a=\frac{q^2}{mv^2+\frac{2 q^2}{r_0}}=\frac{q^2}{3 k_B T+\frac{2 q^2}{r_0}}.
\label{eqA:a}
\end{equation}
Notice that when $T$ is determined, the semi-major axis $a$ is determined. 

The coordinate of the focus (point B) is $(-c,0)$. Assume that the coordinate of point A is $(x_0,y_0)$. We have:
\begin{equation}
\overline{AB}=r_0=\sqrt{(x_0+c)^2+y_0^2}.
\label{eqA:AB}
\end{equation}
Substitute $(x_0,y_0)$ into Eq. (\ref{eqA:hyperbola}) and combine it with Eq. (\ref{eqA:AB}), we obtain:
\begin{equation}
x_0=\frac{a}{c}R,\ y_0=\frac{b}{c}\sqrt{R^2-c^2}.
\label{eqA:A}
\end{equation}
Here, $R\equiv r_0-a$. Using condition $R^2-c^2 \geq 0$, we could easily obtain the relation:
\begin{equation}
b \leq \sqrt{R^2-a^2}.
\label{eqA:b}
\end{equation}
Hence, the maximum of $b$ is $b_\mathrm{m}=\sqrt{R^2-a^2}$.

Differentiate Eq. (\ref{eqA:hyperbola}) and we obtain $\frac{2x}{a^2}\mathrm{d}x - \frac{2y}{b^2}\mathrm{d}y =0$. Hence, the slope of the hyperbola at point A is expressed as:
\begin{equation}
k_0=\tan \theta =\frac{\mathrm{d}y}{\mathrm{d}x} \Big|_{x=x_0}=\frac{b^2}{a^2}\frac{x_0}{y_0}=\frac{b}{a}\frac{R}{\sqrt{R^2-c^2}}.
\label{eqA:k0}
\end{equation}
For one collision, we should emphasize that $\Delta v_y$ is not the change of $v$ in the $y$ direction, but the one in the vertical direction relative to the initial direction of $v$. Hence, the change of $v_y$ can be expressed as $\Delta v_y = v \sin(\pi-2 \theta)$. One collision happens approximately within time $\Delta t =1/ \nu$, and the number of collisions (ions) between $b$ to $b+\mathrm{d}b$ is $ (2\pi b \mathrm{d}b) (v \Delta t) n $. Therefore, the ensemble average of $(\Delta v_y)^2$ can be expressed as:
\begin{equation}
\begin{aligned}
\left<(\Delta v_y)^2\right> &= \int_0^{b_\mathrm{m}} \frac{2\pi b v n}{\nu} \left[v \sin(\pi-2 \theta)\right]^2\ \mathrm{d}b \\
& =\frac{2\pi v^3 n}{\nu} \int_0^{b_\mathrm{m}} b \sin^2 (2 \theta) \mathrm{d}b\\
& = \frac{2\pi v^3 n}{\nu}I.
\label{eqA:dvy2}
\end{aligned}
\end{equation}
Here, the integral term $I=\int_0^{b_\mathrm{m}} b \sin^2 (2 \theta) \mathrm{d}b$, which can be calculated by substituting $\theta$ from Eq. (\ref{eqA:k0}), has been expressed as Eq. (\ref{eq:IT}).

Now, we almost make the approach to the final results. Substitute Eq. (\ref{eqA:Lm}) into (\ref{eqA:dvy2}) and obtain 
\begin{equation}
\left<(\Delta v_y)^2\right> = \frac{\sqrt{2}v^2}{d^2} I.
\label{eqA:dvy22}
\end{equation}
Substitute Eq. (\ref{eqA:dvy22}), (\ref{eqA:Lm}), and $N_s\approx \nu t$ into (\ref{eqA:xvy2}) and obtain:
\begin{equation}
\begin{aligned}
\left<(x v_y)^2\right> &= \nu t\ \frac{2}{3} \frac{v^2}{\nu^2} \ \frac{\sqrt{2} v^2}{d^2} I=\frac{t}{\nu}\frac{2\sqrt{2}v^4}{3d^2}I\\
& =\frac{t}{\sqrt{2}\pi n d^2 v}\frac{2\sqrt{2}v^4}{3d^2}I =t \frac{2v^3}{3\pi n d^4}I.
\label{eqA:xvy22}
\end{aligned}
\end{equation}
At last, substitute Eq. (\ref{eqA:xvy22}) into (\ref{eqA:L}) and then into (\ref{eqA:definition}), and obtain:
\begin{equation}
\begin{aligned}
\eta & = \frac{1}{2t}\frac{V}{k_B T}\ \frac{m^2}{V^2}N\ t \frac{2v^3}{3\pi n d^4}I\\
& = \frac{m^2 v^3}{k_B T\ 3 \pi d^4}I\\
& = \frac{m^2 (3 k_B T/m)^{\frac{3}{2}}}{k_B T\ 3 \pi d^4}I\\
& = \frac{\sqrt{3 m k_B T}}{\pi d^4}I.
\label{eqA:eta}
\end{aligned}
\end{equation}

Finally, we obtain the basic formula of the shear viscosity in our model, which is the same as Eq. (\ref{eq:eta}).
Combining Eqs. (\ref{eq:eta})-(\ref{eq:d}), we obtain the analytical expressions of the shear viscosity for ext-RWSP-VM.
\section*{\label{sec:Ref}References}

\bibliography{RVM_Cheng}
\bibliographystyle{mystyle}

\end{document}